# Quasi-deterministic secure quantum communication using non-maximally entangled states

**Sujan Vijayaraj [1], S. Balakrishnan [2], K. Senthilnathan [2]**

**Abstract**

Quantum communication in general helps deter potential eavesdropping in the course of transmission of bits to enable secure communication between two or more parties. In this paper, we propose a novel quasi-deterministic secure quantum communication scheme using non-maximally entangled states. The proposed scheme follows a simple procedure, and cases where the entanglement required can be significantly reduced to carry out the protocol successfully are discussed. Long sequences or the whole sequence of data can be sent after error checking for a potential eavesdropper. The maximum qubit efficiency of the proposed protocol is found to be 33.333%.

## 1. Introduction

Quantum cryptography exploits the fundamental postulates of quantum mechanics to ensure secure communication between two or more parties. Several quantum communication schemes have been proposed since the introduction of the BB84 protocol in 1984 [1]. In the BB84 protocol, the sender makes use of the rectilinear (R) or diagonal (D) basis to encode information in single photons [2]. The quantum no cloning theorem prevents a malicious third party (Eve) from copying the quantum state during the transmission process. The bases used are publicly announced by both the sender (Alice) and the receiver (Bob). They both reject the bits that were generated through different bases. The remaining bits that were generated through the same basis are preserved. A

Sujan Vijayaraj
sujankvr@gmail.com

S. Balakrishnan
physicsbalki@gmail.com

K. Senthilnathan
senthee@gmail.com

1 School of Electronics Engineering, Vellore Institute of Technology, Vellore, India
2 Department of Physics, School of Advanced Sciences, Vellore Institute of Technology, Vellore, India

subset of these bits are chosen at random and the error rate is calculated, which is expected to be within a certain threshold value. If not, the presence of eavesdropper is disclosed. Alternatively the Ekert protocol relies on the non-locality of a shared maximally entangled pair between the sender and receiver. If the measurement is performed by Alice and Bob in a compatible basis, sifted bits are generated after publicly announcing the bases [3].

Quantum key distribution (QKD) schemes like the BB84 and the Ekert protocol are not used to send bits directly, but instead help establish a private key between the sender and the receiver. After establishing the key, bits are sent through a classical channel using an encryption algorithm. On the other hand, quantum secure direct communication (QSDC) schemes and deterministic secure quantum communication (DSQC) schemes are used to communicate directly over a quantum channel without involving key generation [4]. The first direct communication scheme was proposed by Long and Liu [5]. Over the years, QSDC schemes using single photons, entangled states and superdense coding have been proposed [6-9]. QSDC using single photons was experimentally realized in 2016 [10]. A practical QSDC scheme with a secure communication rate of 50 bps was also implemented [11]. An experimental free space setup was also implemented recently using single photons [12]. DSQC on the other hand uses classical bits to read out the information encoded in the quantum state. DSQC was first proposed by Shimizu and Imoto [13]. A quasi-secure direct communication scheme using EPR pairs, called the ping pong protocol was proposed later on [14]. DSQC using EPR pairs and pure entangled states were proposed in 2006 [15-16]. DSQC using a one time pad [17-18], and higher dimensional quantum systems were also proposed [19].

While most of the proposed schemes use maximally entangled states like Bell states, we have shown how non-maximally entangled states can help leverage security. Moreover, the transmission of maximally entangled states are more difficult than non-maximally entangled states due to decoherence [20]. Hence it is worthwhile to seek quantum communication using non-maximally entangled states. The use of non-maximally entangled states for quantum cryptography has been minimal in spite of their generation in laboratories. The simple act of rotating the polarizer for the pump photons before striking the non-linear crystal in standard parametric down conversion experiments can produce such states. In our work, these states are measured by the R (measurement along 0°, 90° to produce $|0\rangle, |1\rangle$) or D (measurement along 45°, 135° to produce $|+\rangle, |-\rangle$) basis similar to the BB84 protocol, and the measurement result is counted as the bit. Utilizing measuring correlations through entanglement is possible only when the same basis is used. The results 0° or 45° are counted as 0 bit, and 90° or 135° are counted as 1 bit.

The protocol deals with one-way communication from Alice to Bob. Random bits are sent probabilistically using non-maximally entangled states. Two cases for achieving perfect secrecy with reduced entanglement are also discussed. Our protocol uses a one time pad where half of the bits are transmitted using the classical channel. Since the use of non-maximally entangled states introduces imperfections for Bob's bits, it can be referred to as a quasi-DSQC scheme. Unlike most other DSQC schemes, our protocol does not involve the application of any quantum gates, thereby reducing circuit complexity.

The paper is organized as follows. Section 2 describes the steps involved in the protocol. Section 3 elucidates the security of the given protocol. Section 4 is used for discussing the various cases of quantum states and how they affect security. Section 5 concludes the paper.

## 2. The protocol

In this section, we introduce the protocol formally. Let N be the sequence of $n$ number of classical bits to be sent by Alice. The steps involved in the protocol are as follows (see Fig. 1):

1. Alice generates a two qubit non-maximally entangled state $\alpha|00\rangle + \beta|11\rangle$, where $\alpha, \beta \in \mathbb{C}$ and $|\alpha|^2 + |\beta|^2 = 1$.
2. She sends one of the qubits in the entangled pair to Bob, after which they both measure their qubit using the R or D basis randomly.
3. If Bob's measurement outcome is 0, he discerns the bit as 0. If he measures 1, it is a 1 bit.
4. After $n$ number of iterations of transmitting and measuring qubits using Steps 1-3, Bob ends up with a sequence N1 having $n$ bits. Alice and Bob exchange their basis information (whether R or D) through a classical channel. This information can be represented as a sequence with R and D representing 0 and 1 respectively.
5. Bob discards the bits that are generated through different bases while noting their positions. Simultaneously Alice notes the same positions of bits in N, where the bits in these positions were discarded from N1 due to incompatible bases. Let the new sequence of bits derived from these positions in N be A. Also, let the number of bits in the sequence A be $d$. Both Alice and Bob have now generated a separate sequence of sifted bits (bits generated using the same basis) P and Q respectively which will have $n - d$ number of elements. As one can see in the below security analysis, $d \approx n - d \approx n/2$.
6. Bob selects a subset of his sifted bits from Q and sends it along with its respective position to Alice through the classical channel.
7. Alice compares it with the corresponding original bits in P and checks for any error (if the subset of the elements of P and Q are unequal).
8. If the error rate of sifted bits is above a threshold value, Eve's presence is disclosed and the process is terminated at this point. If the error rate is under a threshold value, Alice applies XOR to each bit in A and the corresponding sifted bit, $A \oplus P$ and this result is disclosed to Bob through the classical channel. Let this new bit sequence be G.
9. Bob receives G and applies XOR with his sifted bits, $G \oplus Q = A \oplus P \oplus Q$, which is equivalent to A if P and Q are equal. Bob introduces this result in the position of discarded bits with the previously received sifted bits Q. Hence Bob obtains $n$ bits, of which bits in Q are random.
10. Alice announces the position of those bits in Q that are different from the actual sequence of bits in N (since they are the same as in P). Bob simply complements the corresponding bits in Q. Thus, Bob is left with the final sequence of $n$ bits.

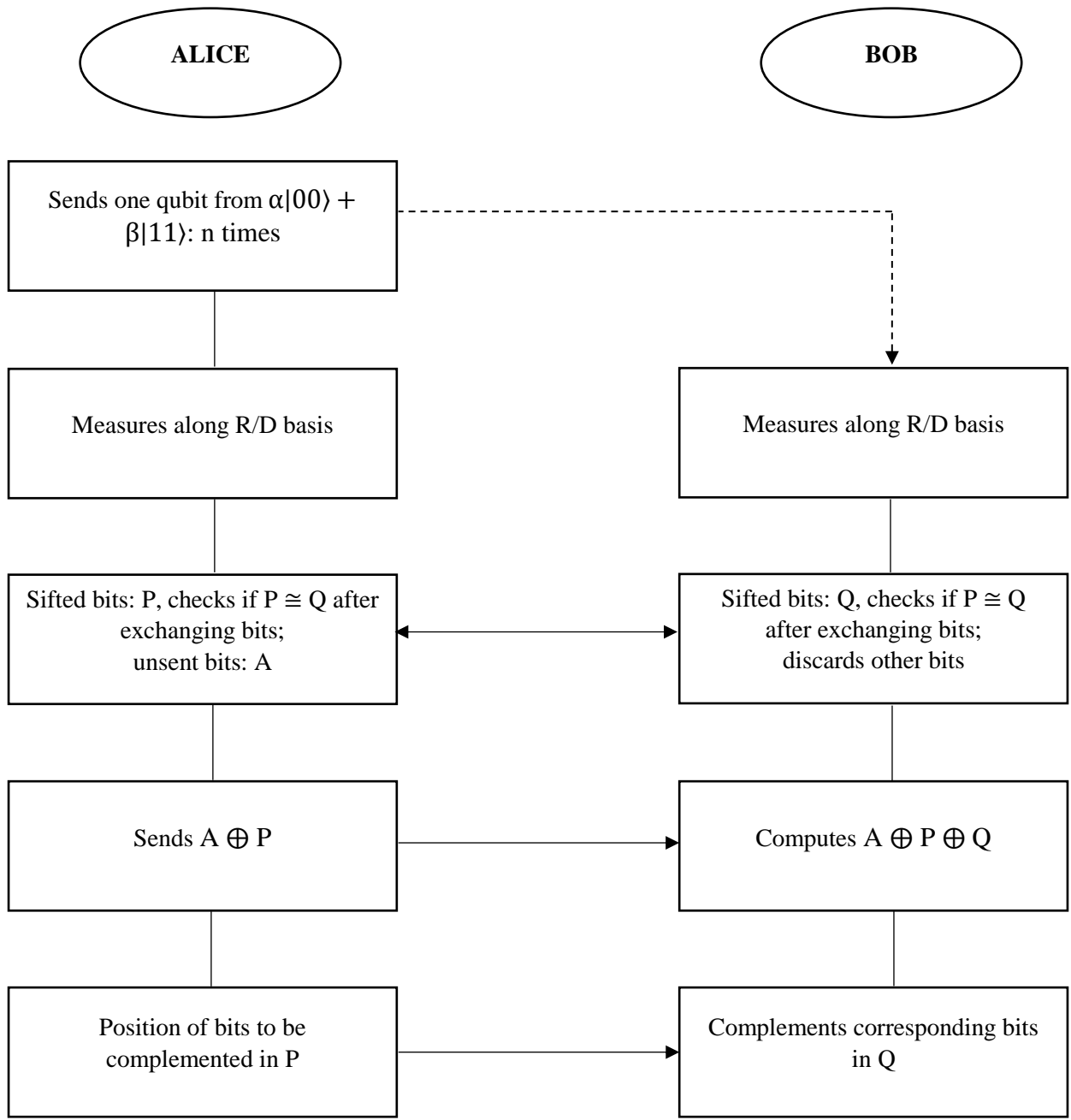

**Fig. 1** Quasi-DSQC protocol between Alice and Bob: dashed arrow indicates quantum channel while the normal ones indicate classical channel

## 3. Security

If Alice and Bob employ the same bases, their measurement outcomes are the same due to the correlation in the entangled state. If Eve intercepts Alice's photons and uses a different basis, her outcome is random. Hence she sends a randomized bit to Bob which might be different from the one that Alice measures from her qubit. This reveals the presence of Eve. If the error rate calculated by Alice in Step 8 exceeds a certain percentage, the presence of Eve can be known. The error rate can be calculated as follows:

Error Rate = Probability of Eve making an error × Probability of Bob making an error

= 50% × 50%

= 25%

The threshold error rate is therefore 25%. Due to incompatible bases, Bob loses 50% of the bits. This can be retrieved by a one time pad using the sifted bits. Alice applies XOR with the bits that Bob lost and the sifted bits. Each of these sequences would have about 50% of the total bits sent by Alice. After Alice sends Bob these new bits, Bob applies XOR with the sifted bits and the new bits that Alice sent. Note that the sifted bits are known only to Alice and Bob. Let *s* be a sifted bit and *f* be a corresponding bit that Bob lost and must receive from Alice. Bob receives $s \oplus f$ from Alice. With that Bob computes

$$s \oplus (s \oplus f) = (s \oplus s) \oplus f = 0 \oplus f = f. \tag{1}$$

Hence Bob receives the previously discarded bits. He has now received all $n$ bits, where some are randomly complemented. If the transmission is free from eavesdropping, Alice can correct this by announcing the position of the wrong bits. Since the sifted bits are obtained randomly, Eve does not obtain any useful information before disclosing her presence. Hence, the proposed protocol offers a solution for communicating against an eavesdropping strategy. Alice can keep a track on the result of Bob's qubit due to the correlation. However, a small percentage of the bits are lost due to the use of non-maximally entangled states. This is explained in the following section.

## 4. Role of various quantum states in security

We use concurrence, $C = 2|\alpha\beta|$, as a measure of entanglement, where $C = 1$ for a maximally entangled state and less than 1 for a non-maximally entangled state ($0 < C \leq 1$) [21]. In the given entangled state in the protocol, substituting for $|0\rangle, |1\rangle$ with $|0\rangle = \frac{1}{\sqrt{2}}(|+\rangle + |-\rangle)$ and $|1\rangle = \frac{1}{\sqrt{2}}(|+\rangle - |-\rangle)$ gives,

$$\alpha|00\rangle + \beta|11\rangle = \frac{1}{2}\{(\alpha + \beta)[|++\rangle + |--\rangle] + (\alpha - \beta)[|+-\rangle + |-+\rangle]\}. \tag{2}$$

This new representation on the RHS must be considered for the case when both Alice and Bob use the D bases for sifting bits. It can easily be seen that the state is not fully correlated when using the D bases. With a small probability, say $P_d$, where $P_d = 0.5|\alpha - \beta|^2 = 0.5(1 - C)$, the state produces valid but different sifted bits while using the D basis. The error among all sifted bits is $0.5P_d$, because there is no such error while using the R bases. This error rate remains the same even after the one time pad transmission. The relation between this bit error and $C$ is summarized in Fig. 2 for decreasing $C$. It can be seen that the number of valid bits transmitted to Bob decreases with decrease in entanglement.

Additionally, classical bits are used in error checking and for the cipher apart from qubits. It is therefore important to quantify the efficiency of the scheme. The theoretical qubit efficiency of a protocol is defined as

$$\eta = \frac{c}{q+b}, \qquad (3)$$

where $c$ is the number of bits received by Bob, $q$ is the number of qubits transmitted by Alice and $b$ is the number of classical bits exchanged between Alice and Bob [22]. Here $c = 1 - 0.5P_d$, $q = 1$ and $b = 1 + 0.5 + 0.5 = 2$, where $b$ includes the basis exchanged, bits from the one time pad and the position of bits to be complemented. Thus $\eta = (2 - P_d)/6 = (3 + C)/12$, which varies between $0.25$ and $0.333$ depending on the entanglement.

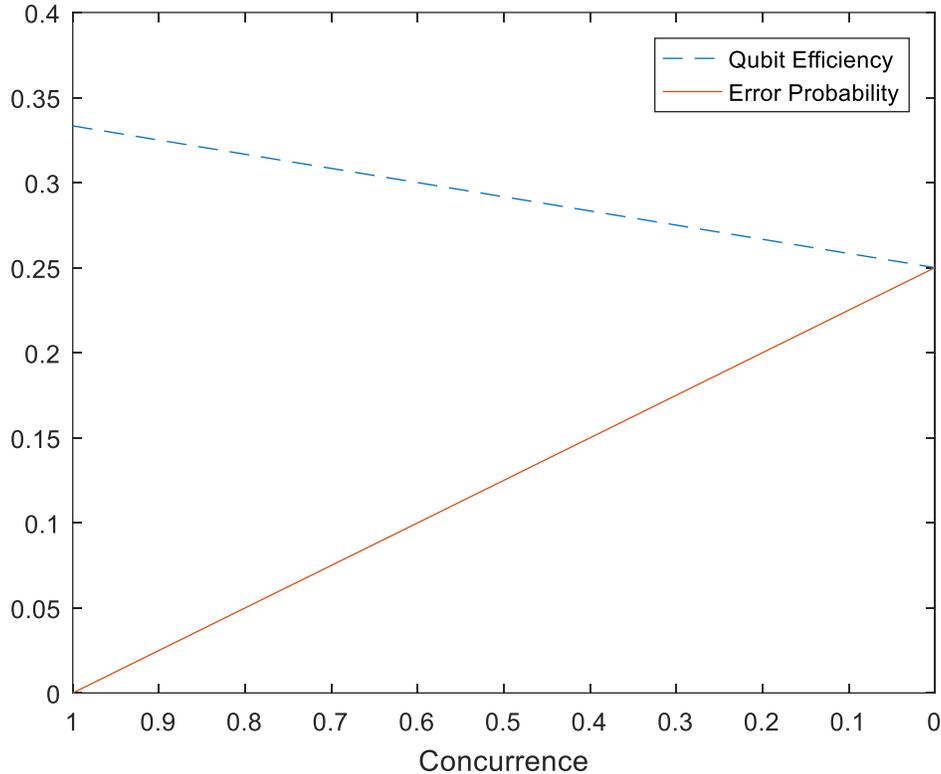

**Fig. 2** Relation between error probability in sifted bits ($0.5P_d$), qubit efficiency with respect to concurrence

If the parties are not satisfied with the protocol not confirming to 100% error free bits, they can sacrifice bits to ensure that it is the case. Let the concurrence be $C_R$ when R basis is used, and $C_D$ when D basis is used. Before using the one time pad, the generated sifted bits can be free of the discussed error within one of the two following cases.

i. When $0 < C_R < 1$ and $C_D = 1$ (average is greater than 0.5), the parties can obtain at most $n$ error free sifted bits from $2n$ entangled states. This is because $P_d = 0$ when $C_D = 1$, causing no error due to sifted bits. Here 50% of the total number of qubits are sifted while also being error free, given that the error rate after checking is within the threshold.

ii. When $0 < C_R \leq 1$ and $0 < C_D < 1$ (average is greater than 0), the parties can obtain at most $n$ error free sifted bits from $4n$ entangled states. This is because the parties must discard all sifted bits generated from the D basis as it has a non-zero probability, $P_d$, of being unreliable. Nevertheless, before discarding them, Alice must use all or a random subset of these sifted bits generated using D basis for error checking. She should ensure that $P_d < 0.5(1 - C_D) < 0.5$, to make sure that Eve does not intercept and resend after measuring solely with the R basis. The usual error checking should also be carried out with a subset of the sifted bits generated from the R basis.

Alice can prepare various entangled states by rotating the pump polarization in her setup with respect to the vertical or horizontal by an angle θ. A general non-orthogonal two qubit entangled state can be expressed as $(\varepsilon|00\rangle + |11\rangle)/\sqrt{\varepsilon^2 + 1}$ where $0 < \varepsilon \leq 1$ and the degree of entanglement, $\varepsilon = \tan\theta$ [23]. Hence, Alice can control α and β by controlling θ to produce different states.

## 5. Conclusion

In the proposed quasi-DSQC scheme, different bases are used along with entangled states to achieve secure communication. The first part of the protocol is similar to the BBM92 QKD protocol [24]. The proposed scheme can also be favorable for transmitting qubits as non-maximally entangled states are more robust in specific decoherence models. Most importantly, it is difficult to retain maximally entangled states during transmission in practical scenarios. One half of the sequence of bits is sent by generating sifted bits and the other half uses a one time pad with the sifted bits. The position of bits to be complemented among the sifted bits is announced by Alice at the end. The scheme also shows how theoretically error free quantum communication can be achieved with non-maximally entangled states at the cost of increased number of entangled states. The resulting qubit efficiency of the protocol is between 25% and 33.333% depending on the concurrence. By handing the responsibility of state preparation to Alice (unlike many other schemes based on entanglement in existing literature), the possibility of Eve introducing entangled states in her favor is removed. However, difficulty with practical realization due to error in transmitting entangled states, such as in the described protocol remains. Quantum repeaters, which can be used to mitigate this, is still a growing line of research in itself [25].

## 6. References

1. Bennett, C. H., Brassard, G.: Quantum cryptography: public key distribution and coin tossing. Theor. Comput. Sci. 560(12), 7-11 (2014)

2. Fox, M.: Quantum optics: an introduction (Vol. 15), pp. 249-252. OUP Oxford (2006)

3. Ekert, A. K.: Quantum cryptography based on Bell's theorem. Phys. Rev. Lett. 67(6), 661 (1991)

4. Long, G. L., Deng, F. G., Wang, C., Li, X. H., Wen, K., Wang, W. Y.: Quantum secure direct communication and deterministic secure quantum communication. Front. Phys. China. 2(3), 251-272 (2007)

5. Long, G. L., Liu, X. S.: Theoretically efficient high-capacity quantum-key-distribution scheme. Phys. Rev. A. 65(3), 032302 (2002)

6. Deng, F. G., Long, G. L.: Secure direct communication with a quantum one-time pad. Phys. Rev. A. 69(5), 052319 (2004)

7. Wang, C., Deng, F. G., Li, Y. S., Liu, X. S., Long, G. L.: Quantum secure direct communication with high-dimension quantum superdense coding. Phys. Rev. A. 71(4), 044305 (2005)

8. Deng, F. G., Li, X. H., Li, C. Y., Zhou, P., Zhou, H. Y.: Quantum secure direct communication network with Einstein–Podolsky–Rosen pairs. Phys. Lett. A. 359(5), 359-365 (2006)

9. Xi-Han, L., Chun-Yan, L., Fu-Guo, D., Ping, Z., Yu-Jie, L., Hong-Yu, Z.: Quantum secure direct communication with quantum encryption based on pure entangled states. Chin. Phys. 16(8), 2149 (2007)

10. Hu, J. Y., Yu, B., Jing, M. Y., Xiao, L. T., Jia, S. T., Qin, G. Q., Long, G. L.: Experimental quantum secure direct communication with single photons. Light Sci. Appl. 5(9), e16144 (2016)

11. Qi, R., Sun, Z., Lin, Z., Niu, P., Hao, W., Song, L., Huang, Q., Gao, J., Yin, L., Long, G.L.: Implementation and security analysis of practical quantum secure direct communication. Light Sci. Appl. 8(1), 1-8 (2019)

12. Pan, D., Lin, Z., Wu, J., Sun, Z., Ruan, D., Yin, L., Long, G.: Experimental free-space quantum secure direct communication and its security analysis. Photon. Res. 8, 1522-1531 (2020)

13. Shimizu, K., Imoto, N.: Communication channels secured from eavesdropping via transmission of photonic Bell states. Phys. Lett. A. 60(1), 157 (1999)

14. Boström, K., Felbinger, T.: Deterministic secure direct communication using entanglement. Phys. Rev. Lett. 89(18), 187902 (2002)

15. Zhu, A. D., Xia, Y., Fan, Q. B., Zhang, S.: Secure direct communication based on secret transmitting order of particles. Phys. Rev. A. 73(2), 022338 (2006)

16. Li, X. H., Deng, F. G., Li, C. Y., Liang, Y. J., Zhou, P., Zhou, H. Y.: Deterministic secure quantum communication without maximally entangled states. arXiv preprint quant-ph/0606007 (2006)

17. Chang, Y., Zhang, S. B., Yan, L. L., Li, J.: Deterministic secure quantum communication and authentication protocol based on three-particle W state and quantum one-time pad. Chinese Sci. Bull. 59, 2835 (2014)

18. Li, N., Li, J., Li, L. L., Wang, Z., Wang, T.: Deterministic secure quantum communication and authentication protocol based on extended GHZ-W state and quantum one-time pad. Int. J. Theor. Phys. 55, 3579 (2016)

19. Jiang, D., Chen, Y., Gu, X., Xie, L., & Chen, L.: Deterministic secure quantum communication using a single d-level system. Sci. Rep. 7, 44934 (2017)

20. Wang, X. W., Tang, S. Q., Yuan, J. B., Kuang, L. M.: Nonmaximally entangled states can be better for quantum correlation distribution and storage. Int. J. Theor. Phys. 54(5), 1461-1469 (2015)

21. Hill, S., Wootters, W. K.: Entanglement of a pair of quantum bits. Phys. Rev. Lett. 78(26), 5022 (1997)

22. Cabello, A.: Quantum key distribution in the Holevo limit. Phys. Rev. Lett. 85(26), 5635 (2000)

23. White, A. G., James, D. F., Eberhard, P. H., Kwiat, P. G.: Nonmaximally entangled states: production, characterization, and utilization. Phys. Rev. Lett. 83(16), 3103 (1999)

24. Bennett, C. H., Brassard, G., & Mermin, N. D.: Quantum cryptography without Bell's theorem. Phys. Rev. Lett. 68(5), 557 (1992)

25. Munro, W. J., Azuma, K., Tamaki, K., & Nemoto, K.: Inside quantum repeaters. IEEE J. Sel. Top. Quantum Electron. 21(3), 78-90 (2015)